\documentclass[final,12pt]{elsarticle}

\usepackage{epsfig}
\usepackage{amsthm}
\usepackage{amsmath} 
\usepackage{amssymb}
\usepackage{amsfonts}
\usepackage{mathpazo}

\usepackage{times}
\usepackage{color}
\usepackage{hyperref} 
\usepackage{tikz}
\usetikzlibrary{automata,arrows,decorations.pathmorphing,decorations.markings,positioning} \usetikzlibrary{shapes} 
\usepackage{scrextend} \addtokomafont{labelinglabel}{\sffamily}
\usepackage{refcount}

\usepackage[textsize=scriptsize]{todonotes}

\newtheorem{theorem}{Theorem}

\newtheorem{lemma}[theorem]{Lemma}

\newtheorem{reptheorem}{Theorem}

\newdefinition{definition}[theorem]{Definition}
\newdefinition{conjecture}[theorem]{Conjecture}
\newdefinition{observation}[theorem]{Observation}
\newdefinition{example}[theorem]{Example} 
\newdefinition{remark}[theorem]{Remark}
\newdefinition{corollary}[theorem]{Corollary}

\newcommand{\N}{{\mathbb{N}}}
\newcommand{\Z}{{\mathbb{Z}}}

\usepackage{amssymb} 
\usepackage{algpseudocode,algorithm} 
\usepackage{subfigure}

\usepackage{algpseudocode,algorithm,algorithmicx}
\newcommand*\Let[2]{\State #1 $\gets$ #2}
\algdef{SE}[DOWHILE]{Do}{doWhile}{\algorithmicdo}[1]{\algorithmicwhile\textsl{}\ #1}%

\renewcommand{\leq}{\leqslant}
\renewcommand{\geq}{\geqslant}

\newcommand{\eqdef}{\triangleq}
\newcommand{\seq}[1]{\left\langle{#1}\right\rangle}
\newcommand{\mathset}[1]{\left\{{#1}\right\}}
\newcommand{\abs}[1]{\left|{#1}\right|}
\newcommand{\cR}{\mathcal{R}}

\newcommand{\head}{\mathsf{head}}
\newcommand{\nxt}{\mathsf{next}}
\newcommand{\tmin}{\tau_{\min}}

\newcommand{\kmo}{k \! - \! 1}
\newcommand{\kmt}{k \! - \! 2}
\newcommand{\spo}{\sigma\! + \! 1}
\newcommand{\sppo}{\sigma'\! + \! 1}

\begin {document}

\sloppy
\journal{Discrete Mathematics}
\begin{frontmatter}
\title{An Efficient Shift Rule for the \\ Prefer-Max De Bruijn Sequence\tnoteref{t1}}

\tnotetext[t1]{This work was supported in part by the Israeli Science Foundation (ISF) under grant no.~130/14, and grant no.~856/16.}

\author[cs]{Gal Amram}
\ead{galamra@cs.bgu.ac.il}

\author[cs]{Yair Ashlagi}
\ead{ashlagi@post.bgu.ac.il}

\author[cs]{Amir Rubin}
\ead{amirrub@cs.bgu.ac.il}

\author[cs]{Yotam Svoray} 
\ead{ysavorai@post.bgu.ac.il}

\author[ee]{Moshe Schwartz}
\ead{schwartz@ee.bgu.ac.il}

\author[cs]{Gera Weiss}
\ead{geraw@cs.bgu.ac.il}

\address[cs]{Department of Computer Science, Ben-Gurion University of The Negev} 
\address[ee]{Department of Electrical and Computer Engineering, Ben-Gurion University of The Negev} 

\begin{abstract}
A shift rule for the prefer-max De Bruijn sequence is formulated, for all sequence orders, and over any finite alphabet. An efficient algorithm for this shift rule is presented, which has linear (in the sequence order) time and memory complexity.
\end{abstract}

\begin{keyword}
De Bruijn sequence \sep Ford sequence \sep prefer-max sequence \sep shift rule
\MSC[2010] 94A55 \sep 05C45 \sep 05C38
\end{keyword}

\end{frontmatter}

\section{Introduction}

A $k$-ary De Bruijn sequence of order $n$ (denoted $(n,k)$-DB), is a word $\seq{\nu_i}_{i=0}^{k^n-1}\eqdef \nu_0 \nu_1 \dots \nu_{k^n-1}$ over the alphabet $[k]\eqdef\mathset{0,1,\dots,k-1}$, i.e., $\nu_i\in[k]$ for all $i\in\Z_{k^n}$, such that all $n$-subwords $\nu_{i} \nu_{i+1}\dots \nu_{i+n-1}$ are distinct (note that $i\in\Z_{k^n}$ means that indices are taken modulo $k^n$).
Of the many $(n,k)$-DB sequences that exist, a particular sequence stands out, featuring in many past works. Consider first the binary case, $k=2$, start the sequence with $0^n$, and add bit by bit, always preferring to append a $1$, unless it creates an $n$-word that has already been seen previously. After obtaining a sequence of length $2^n$, move the $0^n$ prefix to the end of the sequence. The result is an $(n,2)$-DB sequence dubbed the prefer-one sequence or Ford sequence~\cite{Mar34,For57}.
By complementing all the bits, we obtain the prefer-min $(n,2)$-DB sequence. In the non-binary case, we can replace the prefer-one rule by a prefer-max (assuming a lexicographical order of the alphabet), resulting in the lexicographically largest $(n,k)$-DB sequence, and symmetrically (by complementation), a prefer-min $(n,k)$-DB sequence which is the lexicographically smallest $(n,k)$-DB sequence.

The greedy bit-by-bit algorithm of~\cite{Mar34}
is certainly an inefficient way of generating the prefer-max sequence, running in $\Theta(nk^n)$ time (integer operations), and requiring $\Theta(k^n)$ memory. Several suggestions have been made since to improve the efficiency of the sequence construction. Fredricksen and Kessler~\cite{FreKes77}, and Fredricksen and Maiorana~\cite{FreMai78} showed that the prefer-max sequence is in fact a concatenation of certain (Lyndon) words. While seemingly an inefficient way to generate the prefer-max sequence, a later careful analysis~\cite{RusSavWan92} has shown that this decomposition allows us to generate the sequence of length $k^n$ in $O(k^n)$ time.

However, another equally important way of generating sequences is of interest, namely, by using a shift rule. It is well known that any $(n,k)$-DB sequence $\seq{\nu_i}_{i=0}^{k^n-1}$ can be generated by a feedback shift register (FSR) of order $n$, i.e., there exists a shift-rule function $f\colon[k]^n\to [k]$ such that $\nu_{i+n}=f(\nu_i,\nu_{i+1},\dots,\nu_{i+n-1})$ for all $i\in\Z_{k^n}$. Several efficiently computable shift rules for De Bruijn sequences are known, requiring $O(n)$ time and memory to generate the next letter in the sequence, given the preceding $n$ letters (see \cite{Etz86a}, as well as \cite{SawWilWon16} for a comprehensive list). We also mention the recent \cite{DraHerSawWilWon18}, which describes an efficient shift rule for the $k$-ary ``grandmama'' sequence (which is defined by a co-lexicographic order, compared with the lexicographic order of the prefer-max sequence). However, with a single exception, they only generate non prefer-max sequences, and the only exception~\cite{Fre72}, addresses only the generation of binary prefer-one sequences.

The main contribution of this paper is an efficient shift-rule function for the $(n,k)$ prefer-max De Bruijn sequence, $k\geq 2$ (the case of $k=1$ is trivial). The shift rule, given in Algorithm~\ref{alg:pref-max}, is based on the decomposition of the prefer-max sequence found by~\cite{FreMai78}, and operates in $O(n)$ time and memory. This closes a gap in the literature, since while efficient constructions for the entire prefer-max sequence are known, an efficient shift rule is only known in the binary case. 

The paper is organized as follows. In Section~\ref{sec:prelim} we provide the necessary notation used throughout the paper, and recall some relevant results. In Section~\ref{sec:shift} we provide a mathematical expression for the shift rule. We proceed in Section~\ref{sec:alg} to devise an efficient algorithm that implements the shift rule. We conclude in Section~\ref{sec:discuss} with a short discussion of the results.

\section{Preliminaries}
\label{sec:prelim}

Let us start by reviewing the necessary definitions and previous results, before presenting the new results. To avoid trivialities, we assume throughout the paper that $n,k\geq 2$. With our alphabet letters $[k]$ we associate a lexicographical order, $0<1<\dots<k-1$. This order is extended in the natural way to all finite words from $[k]^*$ by defining $x<y$ if either $x$ is a prefix of $y$, or there exist (possibly empty) $u,v,w\in[k]^*$ and two letters $\sigma,\sigma'\in[k]$, $\sigma<\sigma'$, such that $x=u\sigma v$ and $y=u\sigma' w$.

Given a word $v \eqdef \sigma_0 \sigma_1 \dots \sigma_{n-1}$, with $\sigma_i\in[k]$,we define the \emph{rotation} operator, $\cR\colon [k]^n\to [k]^n$ as $\cR(v)\eqdef \sigma_1 \sigma_2\dots \sigma_{n-1} \sigma_0$. We say that two words $v,u\in[k]^n$ are \emph{cyclically equivalent} if there exists $i\in\Z$ such that $v=\cR^i(u)$. The equivalence classes under $\cR$ are called \emph{necklaces}. The number of necklaces, denoted by $Z_k(n)$, is known to be $Z_k(n)\eqdef \frac{1}{n}\sum_{d|n}\phi(d)k^{n/d}$, where $\phi$ is Euler's totient function (also the number of cycles in the pure cycling FSR, see~\cite{LinWil01}). Let $v\in[k]^n$ be a word. The cyclic order of $v$, denoted $o(v)$, is the smallest positive integer $o(v)\in\N$ such that $\cR^{o(v)}(v)=v$ or, alternatively, it is the number of elements in the necklace containing $v$. If $o(v)=\abs{v}$ we say that $v$ is \emph{primitive}. For any $v\in[k]^n$ there is a unique primitive word $w\in[k]^{o(v)}$ such that $v=w^{n/o(v)}$. 

A primitive word that is lexicographically least in its necklace is called a \emph{Lyndon word}. If $L\in[k]^+$ is a Lyndon word, we shall also find it useful to define $L^m$ as an \emph{expanded Lyndon word}\footnote{In some places, by abuse of notation, a lexicographically least representative of a necklace (which coincides with our definition of an expanded Lyndon word) is also called a necklace. We shall not do the same since we shall later require a different representative of a necklace, which might cause a confusion.}, for all $m\in\N$. Additionally, we can arrange all the expanded Lyndon words of length $n$ in increasing lexicographical order
\[L_0^{r_0}<L_1^{r_1}<\dots<L_{Z_k(n)-1}^{r_{Z_{k}(n)-1}},\]
where $r_i\eqdef n/\abs{L_i}$. The main result of~\cite{FreKes77,FreMai78} (rephrased to simplify the presentation) is that the prefer-min $(n,k)$-DB sequence is in fact the concatenation $L_0 L_1 \cdots L_{Z_k(n)-1}$. We shall use this fact later on, and call it the FKM factorization. We also comment that by complementing all the letters via $\psi\colon [k]\to[k]$, $\psi(i)\eqdef k-1-i$, for all $i\in[k]$, the prefer-min $(n,k)$-DB sequence becomes the prefer-max $(n,k)$-DB sequence, and vice versa. We extend $\psi$ to operate on words in the natural way, i.e., applying it to all letters of the word.

\begin{example}
\label{ex:prefmindb}
Fix $n=2$ and $k=3$. We then have the following lexicographical order of expanded Lyndon words,
\[ 00 < 01 < 02 < 11 < 12 < 22\]
hence
\begin{align*}
L_0&=0 & L_1&=01 & L_2&=02 & L_3&=1 & L_4&=12 & L_5&=2,
\end{align*}
and indeed the prefer-min $(2,3)$-DB sequence is $L_0 L_1 L_2 L_3 L_4 L_5 = 001021122$. After complementing each letter we obtain $\psi(L_0 L_1 L_2 L_3 L_4 L_5) = 221201100$, which is the prefer-max $(2,3)$-DB sequence.
\end{example}

\section{Shift-Rule Construction}
\label{sec:shift}
In this section we state and prove our shift-rule construction. For ease of presentation, we work with the prefer-min sequence, while remembering that by simply complementing letters with $\psi$, this is equivalent to working with the prefer-max sequence.

We first require a definition that distinguishes another necklace member that is not necessarily the expanded Lyndon word $L_i^{r_i}$ defined above.

\begin{definition}
A word $v\in[k]^n$ is a necklace \emph{head}, tested by the predicate $\head(v)$, if we can write $v$ as $v=(\kmo)^t w\sigma$, where $w\in[k]^{n-t-1}$, $\sigma\in[k-1]$ (i.e., $\sigma\neq k-1$), and $\cR^t(v)= w \sigma(\kmo)^t$ is an expanded Lyndon word.
\end{definition}

We briefly note that the necklace containing the single word $(\kmo)^n$ does not formally have a necklace head, whereas all other necklaces have a unique necklace head. Additionally, by the above definition, if $(\kmo)^t w\sigma$ is a necklace head, either $w=\varepsilon$ is empty, or it does not start with the letter $k-1$.

We now define our shift rule. Traditionally, a shift rule is a function that takes $n$ consecutive letters in the sequence (i.e., the current state of an FSR generating the sequence) and its output is the next letter. However, we will find it more convenient to define a shift rule as providing the next state of the FSR. Specifically, let $\seq{\nu_i}_{i=0}^{k^n-1}$ be the prefer-min $(n,k)$-DB sequence. A shift rule for the sequence is a function $f\colon [k]^n\to[k]^n$ such that $f(\nu_i \nu_{i+1}\dots \nu_{i+n-1})=\nu_{i+1}\nu_{i+2}\dots \nu_{i+n}$, for all $i\in\Z_{k^n}$.

\begin{definition} 
\label{nxt}
Let $\nxt \colon [k]^n \to [k]^n$ be defined by
\[
\nxt(\sigma w) \eqdef
			\begin{cases}
				w(\spo)   & \text{if $\sigma\neq k-1$ and $\head(w\sigma)$,} \\
				w (\min{S})   & \text{if $\sigma=k-1$ and} \\
                & \text{$S\eqdef\mathset{\sigma'\in[k-1] \colon \head(w\sigma')}\neq\emptyset$,}  \\ 
				w \sigma      & \text{otherwise,}
			\end{cases}
\]
where $\sigma\in[k]$ and $w\in[k]^{n-1}$.
\end{definition}

The main result of this section is the following theorem.

\begin{theorem}
\label{nextisshiftrule}
	$\nxt$ is a shift rule for the prefer-min $(n,k)$-DB sequence.
\end{theorem}

Before proceeding, we provide an example.

\begin{example}
Continuing our running example from Example \ref{ex:prefmindb}, consider again the prefer-min $(2,3)$-DB sequence $001021122$. Take as an example the subword $\sigma w=21$, i.e., $\sigma=2$ and $w=1$. In this case $\nxt(21)$ is computed using the second case of Definition \ref{nxt}, and $S=\mathset{1}$ since $\head(11)$ is true but $\head(10)$ is false. Thus, $\nxt(21)=11$, which is consistent with the sequence.
\end{example}

In order to prove Theorem~\ref{nextisshiftrule} we state a sequence of lemmas, building up to the main result.

\begin{lemma}
\label{el-rotations}
A word $v\in[k]^+$ is an expanded Lyndon word if and only if $v\leq\cR^i(v)$ for all $i\in\Z$ (i.e., it is lexicographically least in its necklace).
\end{lemma}
\begin{proof}
Consider the (unique) decomposition $v=w^t$, where $w$ is primitive. Note that $\cR^i(v)=(\cR^i(w))^t$. Thus, $v\leq\cR^i(v)$ if and only if $w\leq\cR^i(w)$, which holds for all $i\in\Z$ if and only if $w$ is a Lyndon word.
\end{proof}

A first step we take is showing that increasing the rightmost letter that is not $k-1$ in an expanded Lyndon word maintains the expanded Lyndon property.

\begin{lemma}
If $w\sigma(\kmo)^t\in[k]^n$ is an expanded Lyndon word and $\sigma\in[k-1]$ then $w(\spo)(\kmo)^t$ is also an expanded Lyndon word.	
\label{minus-one-is-L}
\end{lemma}
\begin{proof}
If $w(\spo)(\kmo)^t$ starts with $k-1$ then it is equal to $(\kmo)^n$ and the claim follows. Otherwise, write $w(\spo)(\kmo)^t=xy$ and we shall prove that $xy\leq yx$. If $\abs{y}\leq t$ the claim trivially holds. Otherwise, for some word $v$, 
$y=v(\spo)(\kmo)^t$ and $w=xv$. By assumption, $xv\sigma(\kmo)^t\leq v\sigma(\kmo)^tx$. Since $\abs{v}\leq\abs{xv}$, $xv(\spo)(\kmo)^t\leq v(\spo)(\kmo)^tx$ as well.
\end{proof}


We now turn, in the following lemmas, to consider connections between successive expanded Lyndon words, $L_{i}^{r_i}$ and $L_{i+1}^{r_{i+1}}$.

\begin{lemma}
Let $L_i^{r_i} = w\sigma(\kmo)^t\in[k]^n$ be the $i$th expanded Lyndon word in increasing lexicographical order where $\sigma \neq k-1$. Then, the $(i+1)$th expanded Lyndon word, $L_{i+1}^{r_{i+1}}$, is $w(\spo) x$ where $x\in[k]^t$ is the lexicographically smallest word for which $w(\spo) x$ is an expanded Lyndon word.
\label{Lemma:FollowerOfWLi}
\end{lemma}
\begin{proof}
By Lemma~\ref{minus-one-is-L}, $w(\spo)(\kmo)^t$ is an expanded Lyndon word, i.e., $w(\spo)(\kmo)^t=L_j^{r_j}$ for some $j>i$. It then follows that $L_{i+1}^{r_{i+1}}$ must be of the form $w(\spo)x$ as claimed.
\end{proof}

The following lemma combines the shift rule function, $\nxt$, and the lexicographical order of expanded Lyndon words. We use the notation $\nxt^j(\cdot)$, $j\in\N$, to denote the composition of $\nxt$ with itself $j$ times.

\begin{lemma}
\label{lem:nextli}
$\nxt^{\abs{L_i}}(L_i^{r_i})=L_{i+1}^{r_{i+1}}$, for all $i\in[Z_k(n)-2]$.
\end{lemma}

\begin{proof}
Since $i\in[Z_k(n)-2]$, $L_i^{r_i}$ is not the lexicographically last expanded Lyndon word and not the one before it, i.e., 
\begin{align}
\label{eq:notlast}
L_i^{r_i}& \neq (\kmo)^n, & L_i^{r_i}&\neq (\kmt)(\kmo)^{n-1}.
\end{align}
We can therefore write $L_i = w \sigma (\kmo)^t$, $\sigma\in[k-1]$, so $L_i^{r_i} = w \sigma (\kmo)^t L_i^{r_i-1}$. Using these notations
\[\nxt^{\abs{L_i}}(L_i^{r_i})=\nxt^{\abs{w}+1+t}(w\sigma (\kmo)^t L_i^{r_i-1}).\]
Our proof proceeds by establishing the following three facts:
\begin{enumerate}[(a)]
\item $\nxt^{\abs{w}}(w\sigma (\kmo)^t L_i^{r_i-1}) = \sigma (\kmo)^t L_i^{r_i-1} w$ 
\label{item_a}
\item $\nxt(\sigma (\kmo)^t L_i^{r_i-1} w)=(\kmo)^t L_i^{r_i-1} w (\spo)$ 
\label{item_b}
\item $\nxt^{t}((\kmo)^t L_i^{r_i-1} w (\spo))= L_i^{r_i-1} w (\spo) x$, such that $x\in[k]^t$ is the lexicographically smallest word for which $L_i^{r_i-1} w (\spo) x$ is an expanded Lyndon word.
\label{item_c}
\end{enumerate}
Combining the three facts together, we get that $\nxt^{\abs{L_i}}(L_i^{r_i})= L_i^{r_i-1} w (\spo) x$, and by Lemma~\ref{Lemma:FollowerOfWLi}, we get the desired.    
    
We first prove step~(\ref{item_a}). We contend that this step's claim holds since in the first $\abs{w}$ applications of $\nxt$ only the third case of the definition of $\nxt$ (cf.~Definition~\ref{nxt}) takes place. To prove this, we need to show that for any decomposition $w=w_1\tau w_2$, $\tau\in[k]$, $w_1,w_2\in[k]^*$, we have
\[\nxt(\tau w_2 \sigma (\kmo)^t L_i^{r_i-1} w_1)= w_2\sigma (\kmo)^t L_i^{r_i-1} w_1\tau.\]
Hence, we need to show that $w_2 \sigma (\kmo)^t L_i^{r_i-1} w_1\tau$ is not a necklace head, and that if $\tau=k-1$ then there is no $\sigma'\in[k-1]$ such that $w_2 \sigma (\kmo)^t L_i^{r_i-1} w_1 \sigma'$ is a necklace head.

For the first condition, assume to the contrary that the predicate $\head(w_2 \sigma (\kmo)^t L_i^{r_i-1} w_1 \tau)$ is true. By definition, there should exist an integer $0\leq m \leq\abs{w_2}$ such that $w_2 = (\kmo)^m w_3$ and 
\[\cR^m(w_2 \sigma (\kmo)^t L_i^{r_i-1} w_1 \tau)=w_3 \sigma (\kmo)^t L_i^{r_i-1}  w_1 \tau (\kmo)^m\]
is an expanded Lyndon word. However, we note that
\[w_3 \sigma (\kmo)^t L_i^{r_i-1}  w_1 \tau (\kmo)^m=\cR^{\abs{w_1}+1+m}(L_i^{r_i}).\]
Since $0 < \abs{w_1}+1+m < \abs{L_i}$, this contradicts the cyclic order of $L_i^{r_i}$.

As for the second condition, where $\tau = k-1$, assume to the contrary that for some $\sigma'\in[k-1]$, the word $w_2 \sigma (\kmo)^t L_i^{r_i-1} w_1 \sigma'$ is a necklace head. Again, there should exist an integer $0\leq m\leq \abs{w_2}$, such that $w_2 = (\kmo)^m w_3$, and
\begin{equation}
\label{eq:caseb2}
\cR^m(w_2 \sigma (\kmo)^t L_i^{r_i-1} w_1 \sigma')= w_3 \sigma (\kmo)^t L_i^{r_i-1}  w_1 \sigma' (\kmo)^m
\end{equation}
is an expanded Lyndon word. Thus, on the right-hand side of \eqref{eq:caseb2}, the rightmost letter that is not $k-1$, is $\sigma'$. By repeated applications of Lemma~\ref{minus-one-is-L}, we get that we can replace $\sigma'$ by $k-1$ and still have an expanded Lyndon word, i.e., 
\[w_3 \sigma (\kmo)^t L_i^{r_i-1}  w_1 (\kmo)^{m+1} = \cR^{\abs{w_1}+1+m} (L_i^{r_i}) \]
is an expanded Lyndon word. As in the previous case, this contradicts the cyclic order of $L_i$.

The proof of step~(\ref{item_b}) is simpler. We need to show that we fall under the first case in the definition of $\nxt$ (cf.~Definition~\ref{nxt}), i.e., that $(\kmo)^t L_i^{r_i-1} w\sigma$ is a necklace head. That is indeed true since 
\[ \cR^t((\kmo)^t L_i^{r_i-1} w\sigma)=L_i^{r_i-1} w\sigma (\kmo)^t = L_i^{r_i}\]
is an expanded Lyndon word.

Finally, we address step~(\ref{item_c}), where we need to prove that $\nxt^{t}((\kmo)^t L_i^{r_i-1} w (\spo))=L_i^{r_i-1} w (\spo) x$, such that $x\in[k]^t$ is the lexicographically smallest word for which $L_i^{r_i-1} w (\spo) x$ is an expanded Lyndon word. Note that by \eqref{eq:notlast}, $(\kmo)^t L_i^{r_i-1} w (\spo) \neq (\kmo)^n$, so by Lemma~\ref{Lemma:FollowerOfWLi} such an $x$ exists. Additionally, for any $0 \leq i < t$ we have that $\nxt^{i}((\kmo)^t L_i^{r_i-1} {w} (\spo)) = (\kmo)w'$, thus we never fall within the first case of $\nxt$.

Next, we show that for any $0\leq i < t$, $j = t-i-1$, $x = x_1 \tau x_2$, $x_1\in[k]^i$, $\tau\in[k]$, we have that 
\[\nxt((\kmo)^{j+1} L_i^{r_i-1} w(\spo) x_1) = (\kmo)^{j} L_i^{r_i-1} w(\spo) x_1 \tau.\]
We distinguish between two cases depending on $\tau$. For the first case, let $\tau = k-1$. We contend that we do not fall within the second case of $\nxt$. Assume to the contrary that there is some $\sigma'\in[k-1]$ such that $(\kmo)^{j} L_i^{r_i-1} w (\spo) x_1 \sigma'$ is a necklace head. Thus, $L_i^{r_i-1} w (\spo) x_1 \sigma' (\kmo)^{j}$ is an expanded Lyndon word. Looking at its suffix of length $t$, we get
\[ x_1 \sigma' (\kmo)^{j} < x_1 (\kmo) x_2 =  x_1 \tau x_2 = x,\]
which is a contradiction to the minimality of $x$. 

Now, for the case where $\tau\in[k-1]$. By the definition of $x$ we know that $L_i^{r_i-1} w (\spo) x = L_i^{r_i-1} w (\spo)  x_1 \tau x_2$ is an expanded Lyndon word. Using Lemma~\ref{minus-one-is-L}, we get that $L_i^{r_i-1} w (\spo)  x_1 \tau (\kmo)^{j}$ is also an expanded Lyndon word. Therefore, $(\kmo)^{j} L_i^{r_i-1} w (\spo)  x_1 \tau$ is a necklace head. Left to be shown is that 
\[\tau = \tmin \eqdef \min \mathset{ \tau'\in[k] \colon \head((\kmo)^{j} L_i^{r_i-1} w (\spo)  x_1 \tau')}. \]
Assuming to the contrary that $ \tmin < \tau$, then $(\kmo)^{j} L_i^{r_i-1} w (\spo)  x_1 \tmin$ is a necklace head, implying that  $ L_i^{r_i-1} w (\spo)  x_1 \tmin (\kmo)^{j}$ is an expanded Lyndon word. As in the previous case, since 
\[x_1 \tmin (\kmo)^{j} < x_1 \tau x_2 = x,\]
we get a contradiction to the minimality of $x$.  
\end{proof}

Lemma~\ref{lem:nextli} does not apply to the penultimate expanded Lyndon word, for which, by simple inspection of the definition of $\nxt$ we state
\begin{equation}
\label{eq:penult}
\nxt(L_{Z_k(n)-2}^{r_{Z_k(n)-2}})=\nxt((\kmt)(\kmo)^{n-1})=(\kmo)^n.
\end{equation}

We are now in a position to prove the main result.

\begin{proof}[Proof of Theorem~\ref{nextisshiftrule}]
As a first technical step it is easy to verify that $\nxt$ is a shift rule generating some sequence, since indeed for every $\sigma w$, $\sigma\in[k]$, $w\in[k]^{n-1}$, we have $\nxt(\sigma w)= w\sigma'$ for some $\sigma'\in[k]$.

In the next step, let us examine an unknown sequence $\seq{\nu_i}_{i=0}^{k^n-1}$, that is initialized with $\nu_0 \dots \nu_{n-1}=0^n$, and whose following letters are generated using $\nxt$. We define the numbers $s_i\eqdef \sum_{j=0}^{i-1} \abs{L_i}$, for all $0\leq i\leq Z_k(n)-1$. We prove by induction that for all $i\in [Z_k(n)-1]$, $\nu_{s_i} \nu_{s_i + 1}\dots \nu_{s_i + n-1} = L_i^{r_i}$. The proof is immediate since the induction base is our initialization of $\nu_0 \dots \nu_{n-1}=0^n=L_0^{r_0}$, and the induction step is provided by Lemma~\ref{lem:nextli}, since
\[
\nu_{s_{i+1}} \dots \nu_{s_{i+1}+n-1} = \nxt^{\abs{L_i}}(\nu_{s_i}\dots \nu_{s_i+n-1})
= \nxt^{\abs{L_i}}(L_i^{r_i}) = L_{i+1}^{r_{i+1}}.
\]
By this induction, we already have the prefix of the generated sequence to be $L_0 L_1 \dots L_{Z_k(n)-2}$, but we are missing the last Lyndon word, $L_{Z_k(n)-1}=k-1$. This is easily taken care of, since by \eqref{eq:penult},
\begin{align*}
\nu_{s_{Z_k(n)-2}+1}\dots \nu_{s_{Z_k(n)-2}+n} &=\nxt(\nu_{s_{Z_k(n)-2}}\dots \nu_{s_{Z_k(n)-2}+n-1})\\
&=\nxt(L_{Z_k(n)-2}^{r_{Z_k(n)-2}})=\nxt((\kmt)(\kmo)^{n-1})\\
&=(\kmo)^n,
\end{align*}
namely, the last letter is the last Lyndon word,
\[\nu_{s_{Z_k(n)-2}+n}=\nu_{s_{Z_k(n)-1}}=\nu_{k^n-1}=k-1=L_{Z_k(n)-1}.\]
We also observe that the shift rule wraps around the end of the sequence. Indeed, by a simple inspection of Definition~\ref{nxt}, for every $1\leq i\leq n$,
\begin{align*}
\nxt(\nu_{k^n-i}\dots \nu_{k^n-1} \nu_0 \dots \nu_{n-1-i}) &= \nxt((\kmo)^i 0^{n-i})\\
&= (\kmo)^{i-1} 0^{n-i+1} \\
&= \nu_{k^n-i+1}\dots \nu_{k^n-1} \nu_0 \dots \nu_{n-i}.
\end{align*}

As the final step in the proof, by FKM~\cite{FreKes77,FreMai78} this sequence is exactly the prefer-min $(n,k)$-DB sequence.
\end{proof}

We conclude this section by reminding the reader that in order to generate the prefer-max $(n,k)$-DB sequence (instead of the prefer-min one), all that is required is to start the FSR with $(\kmo)^n$, and to use the shift rule $\psi^{-1}\circ\nxt\circ\psi$, where $\psi$ is the complement function defined in Section \ref{sec:prelim}, and $\circ$ denotes function composition.

\section{Efficient Shift-Rule Algorithm}
\label{sec:alg}

Algorithms for implementing shift-rules for the prefer-min (or prefer-max) $(n,k)$-DB sequences are known \cite{Mar34,For57}. These greedy algorithms require $\Theta(k^n)$ memory, and $\Theta(nk^n)$ time in the worst case (since they in fact need to generate the sequence until the position of the desired next letter). The main result of this section is an efficient algorithm, requiring $O(n)$ time and memory, that implements the shift rule we presented in the previous section. By quick inspection, the claim hinges on an efficient implementation of the $\head$ predicate, as well as finding $\min S$ in the second case of $\nxt$.

Our algorithm uses two key components. The first, is the renowned factorization due to Chen, Fox, and Lyndon \cite{CheFoxLyn58}, namely, that every word $w\in[k]^+$ has a unique decomposition $w=w_0 w_1 \dots w_{m-1}$, such that $w_i$ is a Lyndon word for all $0\leq i\leq m-1$, and $w_0\geq w_1\geq \dots \geq w_{m-1}$. We shall call this the CFL factorization of $w$. The second key component is due to Duval \cite{Duv83}, who showed that this unique decomposition may be computed for all $w\in[k]^+$ in $O(\abs{w})$ time and memory.

First, we address the efficiency of computing the predicate $\head$.
\begin{lemma}
\label{lem:effhead}
For any $w\in[k]^n$ it is possible to compute $\head(w)$ in $O(n)$ time and memory.
\end{lemma}

\begin{proof}
Let $i\in\Z$ be the largest integer such that $(\kmo)^i$ is a prefix of $w$. We apply Duval's algorithm~\cite{Duv83} to $\cR^i(w)$ to obtain its CFL factorization $\cR^i(w)=w_0 w_1 \dots w_{m-1}$. Then $\head(w)$ is true if and only if $w_0=w_{m-1}$.
\end{proof}

Next we recall some useful results already known in the literature. A word $w\in\Sigma^*$ is called a \emph{pre-necklace} if there exists $w'\in\Sigma^*$ such that $ww'$ is an expanded Lyndon word. By \cite[Lemma 2.3]{CatRusSawSer00}, a pre-necklace must necessarily be a fractional power of a Lyndon word, i.e., $w=u^m v$, with $u$ being a Lyndon word, $m\geq 1$, and $v$ a proper prefix of $u$. Since the $u^m$ part is a prefix of a CFL decomposition for $w$, this decomposition is unique and it is efficiently computable in $O(\abs{w})$ time and memory. Finally, we recall \cite[Theorem 2.1]{CatRusSawSer00}, whose authors
dubbed the ``fundamental theorem of necklaces''.

\begin{theorem}[Theorem 2.1 of \cite{CatRusSawSer00}]
\label{which-pre}
Let $w=\tau_0 \tau_1 \dots \tau_{n-1}$, with $\tau_i\in[k]$, be a pre-necklace with fractional-power decomposition $w=u^m v$. Then, $w\sigma$, $\sigma\in[k]$, is a pre-necklace if and only if $\sigma\geq \tau_{\abs{v}}$. Furthermore, $w\sigma$ is a Lyndon word if and only if $\sigma > \tau_{\abs{v}}$.
\end{theorem}

We are now in a position to describe the algorithm for $\nxt(\sigma w)$ and prove its correctness.

\begin{algorithm}[H]
	\caption{$\nxt(\sigma w)$.} 
	\label{alg:pref-max}
	\begin{algorithmic}[1]
		\If {$((\sigma<k-1)\wedge \head(w\sigma))$}
        						\label{l:checkcase2}
			\State return $w(\spo)$ \label{l:case1}
       	\ElsIf {$\sigma w=(\kmo)^{n}$}
			\State return $(\kmo)^{n-1}0$
            					\label{l:case2special}
		\ElsIf  {$((\sigma=k-1)\wedge \head(w(\kmt)))$}
        						\label{l:if}
            \State let $w'=\tau_0\dots \tau_{n-t-1}$, such that $\tau_i\in[k]$, $\tau_0\neq k-1$, $w=(\kmo)^tw'$
			\State let $u^m v=w'$ be the fractional-power decomposition of $w'$\label{line:sesqui}
 			\Let {$\sigma'$}{$\tau_{\abs{v}}$}  \label{line:vth}
			\If {$\head(w\sigma')$}
				\State return $w\sigma'$ \label{line:ret1}
			\Else
				\State return $w(\sppo)$ \label{line:ret2}
        	\EndIf \label{l:endif}
        \Else 
            \State return $w\sigma$ \label{l:case3}
        \EndIf 
	\end{algorithmic}
\end{algorithm}


\begin{theorem}
\label{th:mainalg}
Algorithm \ref{alg:pref-max} correctly computes the shift rule $\nxt$ from Definition~\ref{nxt} in $O(n)$ time and memory.
\end{theorem}

\begin{proof}
We argue that Algorithm~\ref{alg:pref-max} computes the function $\nxt$. We consider the three cases of Definition~\ref{nxt} separately. First, if $\sigma\in[k-1]$ and $\head(w\sigma)$, the algorithm returns $w(\spo)$ in line~\ref{l:case1} as required by the first case of Definition~\ref{nxt}.

Now, assume the input $\sigma w$ falls within the third case of Definition~\ref{nxt}. If $\sigma<k-1$ the claim is obvious as the condition in line~\ref{l:checkcase2} does not hold. If $\sigma=k-1$, then we have $S\eqdef\mathset{\sigma''\in[k-1] \colon \head(w\sigma'')}$. By Lemma~\ref{minus-one-is-L}, $S\neq\emptyset$ if and only if $\head(w(\kmt))$ holds. Thus, line~\ref{l:if} correctly checks whether the second case of Definition~\ref{nxt} applies. We therefore reach line~\ref{l:case3} exactly when the third case of Definition~\ref{nxt} applies, and correctly return $w\sigma$.

We are left with the second case of Definition~\ref{nxt}, where $\sigma=k-1$ and $S\neq\emptyset$. First, the special case of $\sigma w=(\kmo)^n$, is handled correctly in line~\ref{l:case2special}. Otherwise, $w$ contains some letter other than $k-1$, and $w'$ is well defined.

We now contend that $\min S\in\mathset{\sigma',\sigma'+1}$. Since $\head(w(\kmt))$ holds, then $w'(\kmt)(\kmo)^t$ is an expanded Lyndon word, hence $w'$ is a pre-necklace. Also, note that if $\sigma''\in S$ then $w'\sigma''$ is a pre-necklace. By Theorem~\ref{which-pre}, if $\sigma''<\sigma'$ then $w'\sigma''$ is not a pre-necklace. Hence, $\min S\geq \sigma'$. However, also by Theorem~\ref{which-pre}, $w(\sppo)$ is a Lyndon word, thus $\sigma'+1\in S$ and $\min S\leq \sigma'+1$. This leaves only two possible values for $\min S$, and consequently, the algorithm terminates in line \ref{line:ret1} or in line \ref{line:ret2}, and returns the desired word.

Finally, as already noted, CFL factorization, $\head$, as well as the fractional-power decomposition of line \ref{line:sesqui}, may be computed in linear time and memory (all relying on the CFL factorization algorithm). Thus, the entire algorithm takes linear time and memory.
\end{proof}

\section{Discussion}
\label{sec:discuss}

In this paper we studied the well known prefer-min and prefer-max $(n,k)$-DB sequences. We completed a gap in the literature by presenting a shift-rule for the sequences, as well as an efficient algorithm computing this shift rule. The algorithm receives as input a sub-sequence of $n$ letters, and determines the next letter in $O(n)$ time and memory.

The shift rule we presented may be seen as an extension to the binary shift rule presented in \cite{Fre72}. Indeed, if we set $k=2$ in our algorithm, the second case of Definition~\ref{nxt} becomes degenerate, we are left with the algorithm of~\cite{Fre72}. This also explains the main difficulty in our solution, which is finding $\min S$ efficiently. The crux of solving this difficulty is the proof that we only need to choose between two carefully chosen values.

\bibliographystyle{elsarticle-num}
\bibliography{allbib} 

\end{document}